\documentclass[aps,prl,superscriptaddress,reprint,amsfonts,amsmath]{revtex4-1}
\usepackage[final]{graphicx}

\let\oigr\includegraphics
\def\includegraphics[#1]#2{\IfFileExists{#2.eps}{\oigr[#1]{#2}}{\oigr[#1]{figures/#2}}}
\usepackage[usenames,dvipsnames]{xcolor}

\usepackage{comment}

\renewcommand{\vec}[1]{\mathbf{#1}}

\newcommand{\subref}[2]{\hyperref[#1]{#2}}

\usepackage[normalem]{ulem}

\usepackage{hyperref}
\usepackage[usenames,dvipsnames]{xcolor}

\usepackage{tikz}
\hypersetup{colorlinks=true, linkcolor=BrickRed, urlcolor=blue!50!black, citecolor=blue!50!black}

\renewcommand{\vec}[1]{\mathbf{#1}}
\renewcommand{\phi}[0]{\varphi}

\usepackage[normalem]{ulem}

\usepackage[usenames,dvipsnames]{xcolor}
\usepackage[normalem]{ulem}
\newcommand\TODO[1]{%
  {\fboxsep=.3ex\colorbox{BrickRed}{\textcolor{white}{\textsc{TODO}}}
  \textcolor{BrickRed}{#1}}%
}

\begin{document}
%\title{Independence of the Kinetic Glass Transition on Kinetic Parameters}
%\title{How to Obstruct Glassy Relaxation by Increasing Mobility}
\title{How Glassy Relaxation Slows Down by Increasing Mobility}
\date{\today}
\def\dlr{\affiliation{Institut f\"ur Materialphysik im Weltraum,
  Deutsches Zentrum f\"ur Luft- und Raumfahrt (DLR), 51170 K\"oln,
  Germany}}
\def\hhu{\affiliation{Department of Physics,
  Heinrich-Heine Universit\"at D\"usseldorf,
  Universit\"atsstr.~1, 40225 D\"usseldorf, Germany}}
\def\inn{\affiliation{Institut f\"ur Theoretische Physik,
  Universit\"at Innsbruck, Austria}}

\author{Suvendu Mandal}\inn
\author{Thomas Franosch}\inn
\author{Thomas Voigtmann}\dlr\hhu

\begin{abstract}
We investigate how structural relaxation in mixtures with strong dynamical
asymmetry is affected by the microscopic dynamics.
Brownian and Newtonian dynamics simulations of
dense mixtures of fast and slow hard spheres
reveal a striking trend reversal.
Below a critical density, increasing the mobility of the fast particles
fluidizes the system,
yet, above that critical density,
the same increase in mobility strongly hinders
the relaxation of the slow particles.
The critical density itself does not depend
on the dynamical asymmetry and can be identified with the
glass-transition density of the mode-coupling theory.
The asymptotic dynamics close to the critical density is universal,
but strong pre-asymptotic effects prevail in mixtures
with additional size asymmetry.
This observation reconciles earlier findings of a strong dependence on kinetic parameters
of glassy dynamics in
colloid--polymer mixtures
with the paradigm that the glass transition is determined
by the properties of configuration space alone.
\end{abstract}

\maketitle

The microscopic description of classical many-particle systems
has different starting points depending on whether one models a
small-molecular system or a colloidal suspension. Molecular dynamics
follows Newton's equations of motion, while the coarse-grained description
of the solvent in colloidal suspensions leads to stochastic Brownian
dynamics of the colloids.
Since in thermodynamic equilibrium of classical mechanics
the kinetic degrees of freedom can be integrated out separately,
the \emph{equilibrium} phase behavior of colloids is the same as that
of an atomic system, provided the effective interactions among the particles
are the same
\cite{Terentjev.2015,colloids:Dhont.1996}.
This has inspired the use of ``colloids as big atoms''
for all structural and not inherently kinetic features \cite{colloids:Poon.2016}.
In essence, the description reduces from one in phase space to one in
the much smaller configuration space.

%This equivalence needs not hold for the description of time-dependent features.
The glass transition is a well-known \emph{kinetic} phenomenon.
The near-equilibrium long-time dynamics close to the
glass transition shares many
generic features among colloidal and molecular glass-formers
\cite{Hunter:2012Review,rheology:Bouchbinder.2011,glass-simulation:Gleim.1998,glass-theory:Szamel.2004c,glass-theory:Goetze.2009,colloidal-glass:Weeks.2002,colloidal-glass:Pusey.1987,glass-theory:Szamel.1991},
although this is not at all obvious and requires
explanation. The dynamics of
a tracer particle in a random heterogeneous medium (the Lorentz-gas model)
close to the localization transition provides a counter-example: There,
Newtonian and Brownian systems show different dynamical critical exponents
and hence belong to different universality classes \cite{Halperin_PRL:1985,Machta:1986,stat-ph:Spanner.2016}.
%The unviersality classes split because the narrow gaps between frozen-in
%obstacles are probed differently depending on the dynamics of the mobile tracer.

Mixtures of species with high ``dynamical asymmetry'' are a more realistic model
system for cases where the heterogeneous environment is slowly evolving
over time, e.g., in molecular crowding in cells
\cite{biophys:Hoefling.2013,ISI:000327244200007,ISI:000363867400011}.
Relaxation phenomena in such mixtures share features of both the glass transition and
the localization transition
\cite{glass-confined:Krakoviack2005,glass-mixtures:Moreno2006b,cond-mat:Kurzidim.2009,glass-mixtures:Kim.2009,glass-mixtures:Voigtmann.2009,stat-ph:Skinner.2013,glass-theory:Sentjabrskaja.2016}.
A striking
dependence on the mass ratio of the species
was reported for a mixture of large and
penetrable small particles (Asakura-Oosawa model)
\cite{glass-mixtures:Zaccarelli.2004}:
Upon addition of equally heavy small
particles (mass ratio $\alpha_m=m_\text{large}/m_\text{small}=1$) structural relaxation slowed down,
while the addition of lighter small particles ($\alpha_m=10000$)
caused the dynamics to speed up.

If one accepts that such a mass-ratio dependence influences the glass transition
point, this challenges fundamentally
our understanding of the glass transition, because it would entail that
in describing slow dynamics
the reduction of phase space to configuration space
is not admissible. The use, e.g, of colloidal model
systems to understand molecular glasses, or the extrapolation of molecular-dynamics-simulation results
to soft-matter systems would be questionable.
One would also contest the validity of the mode-coupling theory
of the glass transition (MCT), where any
mass-ratio dependence explicitly drops out of the equations that determine the
glass-transition point and the dynamics asymptotically close to it
\cite{glass-theory:Goetze.2003,glass-mixtures:Foffi.2003,glass-theory:Goetze.1987b,glass-theory:Goetze.2009}.
Generally, the great success of classical configuration-space statistical
physics approaches close to the glass transition
\cite{Hunter:2012Review,Berthier:2011Review,Ediger:2012Perspective}
would then appear puzzling.

As we demonstrate here, the
resolution lies in the fact that one has to carefully distinguish
\emph{asymptotic} from \emph{pre-asymptotic} features in the dynamics.
Regarding the glass-transition point itself, ``kinetic universality'' is
found, i.e., the modes of short-time relaxation become irrelevant in determining
the ultimate fate of the system (fluid or glassy).
In the vicinity of this transition point,
two qualitatively different trends emerge upon changing the kinetic parameters.
The MCT glass transition separates a regime where mobile species cause
the relaxation to speed up, from one where they slow down the overall
dynamics.
Studying carefully the influence of dynamical asymmetry thus provides a
way to identify separate mechanisms of structural relaxation.

We start with a simple model, viz.\ a mixture of equal-sized
Brownian hard spheres with different short-time diffusivities.
A similar system with continuous interactions
has recently been studied~\cite{cond-mat:Grosberg.2015,PRL:Weber.2016}
as a model for motility-induced phase separation of active particles.
%Mixtures of ``hot'' and ``cold'' particles with stiff harmonic
%interactions were found to phase separate for sufficiently large ratio
%of short-time diffusivities; again a striking deviation from the
%expectations of configuration-space statistical physics.
%
The use of true hard-sphere particles provides an important conceptual
simplification. For Brownian hard spheres, interparticle forces vanish
for all admissible configurations and only provide no-flux boundary
conditions for the $N$-particle probability density in the case that
two spheres touch. As a consequence, the uniform equilibrium distribution
is still a stationary solution of the Smoluchowski equation, and,
by virtue of the $H$-theorem, the only one. Thus, all equilibrium
properties of the system remain unchanged by assigning fast and slow
short-time diffusivities to the particles.
(This is not necessarily the case if one includes non-equilibrium driving
forces as in active Brownian particles~\cite{Tanaka.2016prf}.)

We compare our Brownian-dynamics (BD) simulations of mixtures of hard-sphere particles with different short-time diffusivities to Newtonian molecular-dynamics (MD)
simulations with different mass ratios.
Particles are slightly polydisperse ($10\%$; $N=1000$ particles
drawn from a Gaussian distribution).
True hard-sphere interactions are guaranteed by an event-driven
algorithm both for Newtonian and Brownian dynamics \cite{comp-sim:Scala.2007,event-drivenMD:Rapaport2004}.
We study the dependence on the short-time kinetic parameters, i.e.,
the ratio of short-time diffusivities $D^0_\text{fast}/D^0_\text{slow}
=:\alpha\ge1$ for BD, and the mass ratio
$m_\text{large}/m_\text{small}=:\alpha_m\ge1$ for MD.
Both parameters are defined such that large $\alpha$ or $\alpha_m$ implies
that the small and fast species is much more mobile than the other species, and
for simplicity, we restrict to the case were small is fast.
MCT calculations are performed based on the Percus-Yevick static structure
factor, as outlined in Refs.~\cite{glass-theory:Goetze.2003,glass-theory:Voigtmann.2011}.
Additional MD simulations of Asakura-Oosawa mixtures are performed to extend
those of Ref.~\cite{glass-mixtures:Zaccarelli.2004} (with $N_\text{large}=500$ large particles of $10\%$ polydispersity, at size ratio $\delta=0.15$).

As a simple quantity that highlights the dynamical behavior, we discuss
the mean-squared displacements (MSD),
$\delta r_a^2(t)=\langle|\vec r_a(t)-\vec r_a(0)|^2\rangle$,
where $a$ labels the species (fast or slow).
For short times, the MSD in BD reflect the dynamical asymmetry as
$\delta r_a^2(t)\simeq 6D^0_at$ and thus $\delta r_\text{fast}^2(t)/\delta r_\text{slow}^2(t)\simeq\alpha$ for $t\to0$.

\begin{figure}
\includegraphics[width=\linewidth]{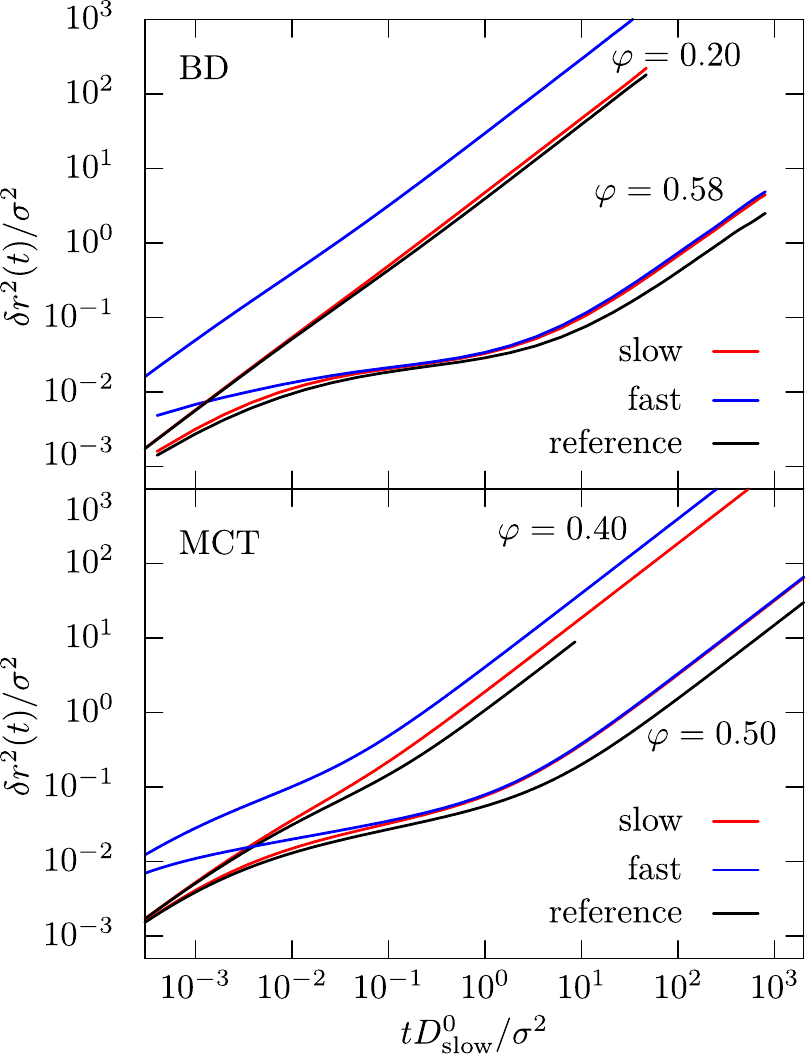}
\caption{
Mean-squared displacement of an equimolar mixture of Brownian hard spheres (diameter $\sigma$) with dynamical asymmetry (ratio of short-time diffusivities) $\alpha=D^0_\text{fast}/D^0_\text{slow}=10$; shown in units of the slow-particle free diffusion time. The MSD for fast (slow) particles is shown as blue (red) lines. Black lines indicate the MSD for $\alpha=1$ for reference. Top panel: BD simulations at $\varphi=0.2$ and $\varphi=0.58$ (left to right). Bottom panel: MCT results for $\varphi=0.4$ and $0.5$ (left to right).
}
\label{fig:msd}
\end{figure}

Figure~\ref{fig:msd} shows the MSD for a $50:50$ mixture of equal-sized
Brownian hard spheres with a short-time diffusivity ratio $\alpha=10$.
All curves show the expected approach to the kinetic arrest transition with
increasing packing fraction $\varphi$: after the short-time diffusive regime,
the MSD show subdiffusive behavior on a length scale comparable to the
typical localization length of particles in the glass, known as the Lindemann
length scale, $r_\text{loc}\approx 0.1\sigma$. At larger length scales,
$\delta r_a^2(t)/\sigma^2\gtrsim1$,
the MSD follow a long-time diffusive asymptote, $\delta r_a^2(t)\simeq 6D_a^Lt$ with long-time diffusion coefficient $D_a^L$,
on a time scale that diverges
upon approaching the glass transition.

For the BD simulation results (top panel of Fig.~\ref{fig:msd}), the case
$\varphi=0.20$ represents the low-density regime where particles interact
only weakly, and $D_a^L\approx D^0_a$. Close inspection reveals that in comparison
to the monodisperse case (all short-time diffusion coefficients equal),
the long-time dynamics of the slow particles is slightly enhanced while
that of the fast particles is slowed down.
In essence, the slow particles play the role of obstacles that hinder the
motion of the fast particles, and the fast particles serve to increase
the effective thermal noise that gives rise to the mobility of the slow
particles at long times.

This trend prevails for densities close to the glass transition, as
exemplified for $\varphi=0.58$ for the BD simulations. There, however,
the fast particles are slowed down much more strongly. While initially,
fast-particle diffusion is faster than that of the slow particles
by a factor $\alpha=10$, their long-time diffusivity is faster only by
less than $20\%$.

Numerical results from MCT (bottom panel of Fig.~\ref{fig:msd}) are in
qualitative agreement with the simulation. As anticipated in the theory,
at high densities the transient nearest-neighbor cageing of particles is the
dominant mechanism of structural relaxation that slows down the dynamics.
Cageing corresponds to
an intermediate-time plateau in the MSD, $\delta r_a^2\approx 6r_\text{loc}^2$.
Since both species share the same excluded-volume interactions,
$r_\text{loc}\approx0.06\sigma$ is independent of the species. Thus,
in the intermediate-time window, the
MSD of both fast and small species become identical.

\begin{figure}
\includegraphics[width=\linewidth]{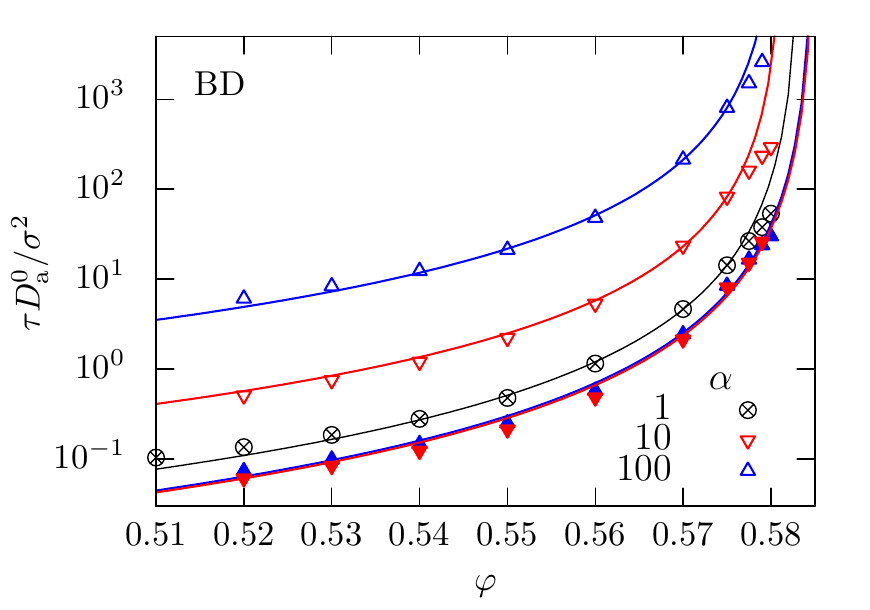}
\caption{
  Structural relaxation times $\tau$ of the self-intermediate
  scattering function to wave number $q\sigma = 7.0$ of the slow
  (filled symbols) and fast (open symbols) partices of an equimolar
  Brownian hard-sphere
  mixture with dynamical asymmetries $\alpha=D^0_\text{fast}/D^0_\text{slow}$,
  as a function of
  packing fraction $\varphi$ and in units of the respective
  short-time diffusion coefficients.
  Symbols are BD simulation results, lines are corresponding power-law
  fits $A(\varphi_c-\varphi)^{-\gamma}$ with $\varphi_c=0.583$ and
  $\gamma=2.3$.
}
\label{fig:tau}
\end{figure}

To demonstrate that a change in $\alpha$ does not change
the MCT glass-transition point, we
show in Fig.~\ref{fig:tau} the  structural relaxation times $\tau$,
extracted from the BD simulations
as the point where the tagged-particle density correlation
function for a wave number close to the static-structure peak has decayed to $1/e$.
The relaxation times are shown in natural units associated with the
short-time motion of each of the species. This highlights the relative
speeding up of slow particles by increasing the dynamical size asymmetry,
and the associated slowing down of fast particles.
For the fast particles, the effect is more drastic: for $\alpha=100$, the
relaxation time of the fast-particle density fluctuations increases by up
to a factor of $50$, while the slow-particle dynamics speeds up
by less than a factor of $2$.

Asymptotically close to the glass transition, MCT predicts a power-law
divergence, $\tau\sim|\varphi-\varphi_c|^{-\gamma}$. Both the critical
point $\varphi_c$ and the non-universal exponent
$\gamma$ are determined solely by the static structural quantities
of the system \cite{glass-theory:Goetze.2009} and thus do not depend on $\alpha$.
In the window of packing fractions
$\varphi\in[0.53,0.575]$, the $\tau(\varphi)$ curves from BD are well described
by such power laws. All data sets can be consistently
rectified using
$\varphi_c=0.583\pm0.002$ and $\gamma=2.3\pm0.2$.
These values of $\varphi_c$ and $\gamma$ are consistent with those found for
nearly-monodisperse hard-sphere systems~\cite{glass-theory:Goetze.2009}.
Thus, both the thermodynamic equilibrium phase diagram and the
kinetic MCT-glass transition in a mixture of hard spheres with different
short-time kinetics are independent of the dynamical size asymmetry.

\begin{figure}
\includegraphics[width=\linewidth]{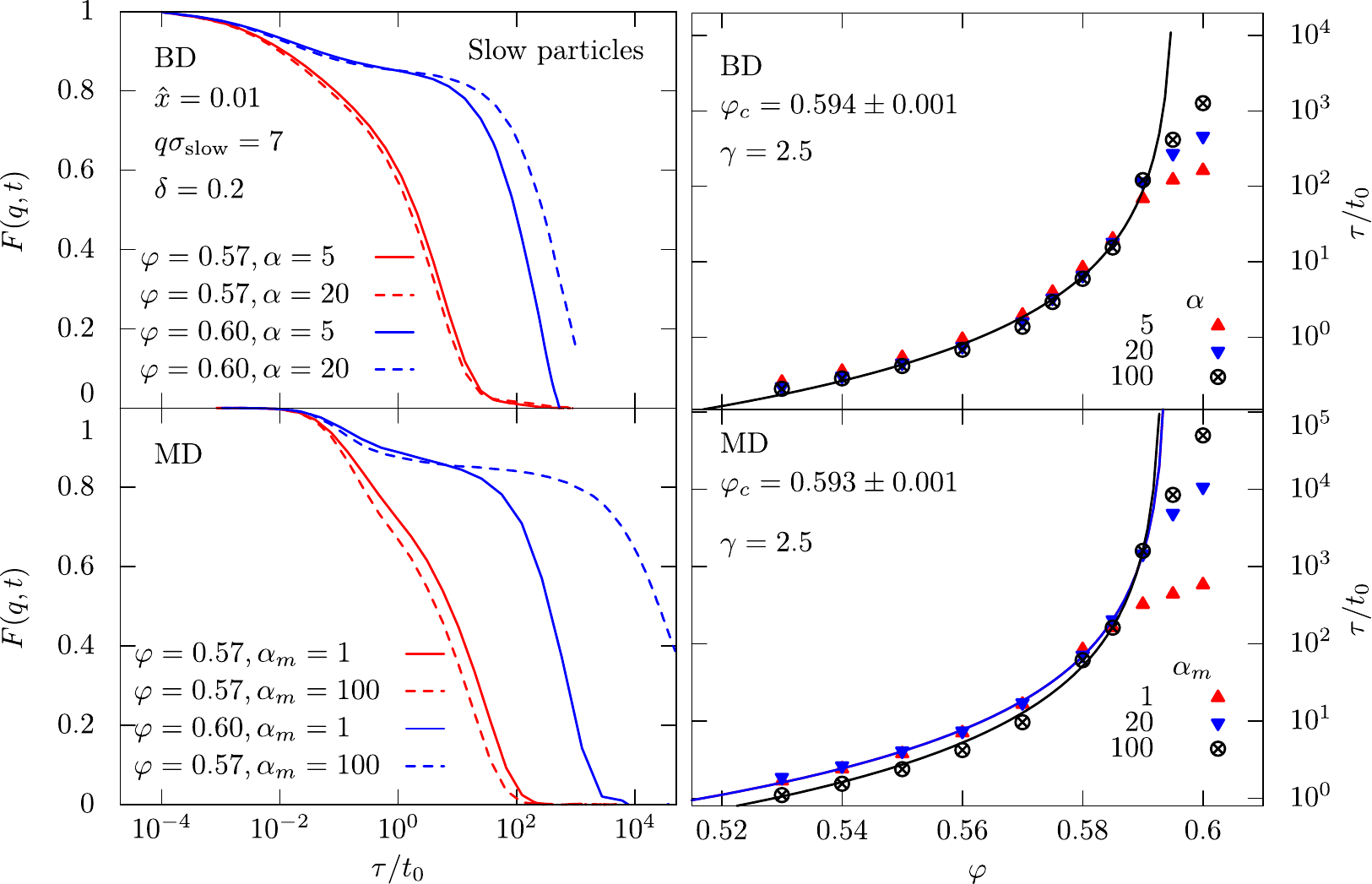}
\caption{Left: Self-intermediate scattering functions (wave number
  $q\sigma_\text{slow}=7.0$) of the
  slow particles in binary hard-sphere mixtures with size ratio
  $\delta=\sigma_\text{fast}/\sigma_\text{slow}=0.2$ and relative
  packing contribution $\hat x=\varphi_\text{small}/\varphi=0.01$. Data are
  shown for
  total packing fractions $\varphi=0.57$ and $0.60$, for
  BD simulations (top panel, $t_0=\sigma^2_\text{slow}/D^0_\text{slow}$)
  with short-time diffusivity ratios $\alpha=5$ (solid lines)
  and $\alpha=20$ (dashed), and for
  MD simulations (bottom panel, $t_0=\sqrt{m\sigma^2_\text{slow}/k_BT}$)
  with mass ratios $\alpha_m=1$ (solid) and $\alpha_m=100$ (dashed).
  Right: structural relaxation times for various $\alpha$ and $\alpha_m$
  (symbols), and power-law fits (lines).
}
\label{fig:binmix}
\end{figure}

%At $\varphi\approx0.58$, deviations from the power-law fits are seen; they
%reflect a well-known deficiency of MCT in neglecting certain relaxation
%processes \cite{glass-theory:Goetze.2009,glass-theory:Szamel.2004c}.
%We will return to a discussion of this residual dynamics in the context
%of size-asymmetric binary mixtures below.

At densities close to and above $\varphi_c$, deviations from the
scenario described above set in. To see this more clearly,
we turn to mixtures that include in addition to dynamical asymmetry also a
strong size asymmetry $\delta=\sigma_\text{fast}/\sigma_\text{slow}\ll1$ in the steric interactions.
For packing fractions $\varphi\lesssim\varphi_c$, the effect of increasing
the dynamical asymmetry is qualitatively as discussed above:
increasing $\alpha$, the structural relaxation of
the slow particles speeds up slightly, see Fig.~\ref{fig:binmix}.
This also holds true in MD upon increasing
$\alpha_m$.
Power-law fits to the structural relaxation times
work in the regime
$\varphi\lesssim0.585$, as demonstrated in the lower panels of Fig.~\ref{fig:binmix}. They yield values $\varphi_c\approx0.594$ and $\gamma\approx2.5$
that do not depend on $\alpha$ or $\alpha_m$: they
confirm that the glass transition point and the asymptotic approach to
it depend neither on the dynamical asymmetry nor on the fact whether the
short-time motion is Brownian or Newtonian \cite{glass-simulation:Gleim.1998,glass-theory:Szamel.2004c}.

A crucial observation is that the MCT transition point divides the
state space into two regimes with different dependence on the kinetic
parameters: Below $\varphi_c$, the trend just described prevails, i.e.,
$\tau$ decreases with increasing $\alpha$ or $\alpha_m$. Yet,
in the high-density regime $\varphi\gtrsim\varphi_c$, this trend is
reversed: here, the structural relaxation of the large particles slows down
upon increasing the dynamical asymmetry.
This is also clearly seen by a shift of structural relaxation to longer times
in the density correlation functions for
$\varphi=0.60$ (left panels of Fig.~\ref{fig:binmix}).

The trend reversal in the dependence on $\alpha$ or $\alpha_m$
separates the dynamical regime of MCT from a regime of relaxation within the
glass, even without the need to precisely determine $\varphi_c$.
The dynamical window over which MCT provides an accurate description of
the data increases with increasing dynamical asymmetry, as is evident
from the right panels of Fig.~\ref{fig:binmix}.
This implies that for $\alpha$ or $\alpha_m$ close to unity, a determination
of the MCT transition point $\varphi_c$ using only power-law fits to
the relaxation times (or similar quantities) becomes more difficult.

\begin{figure}
\includegraphics[width=\linewidth]{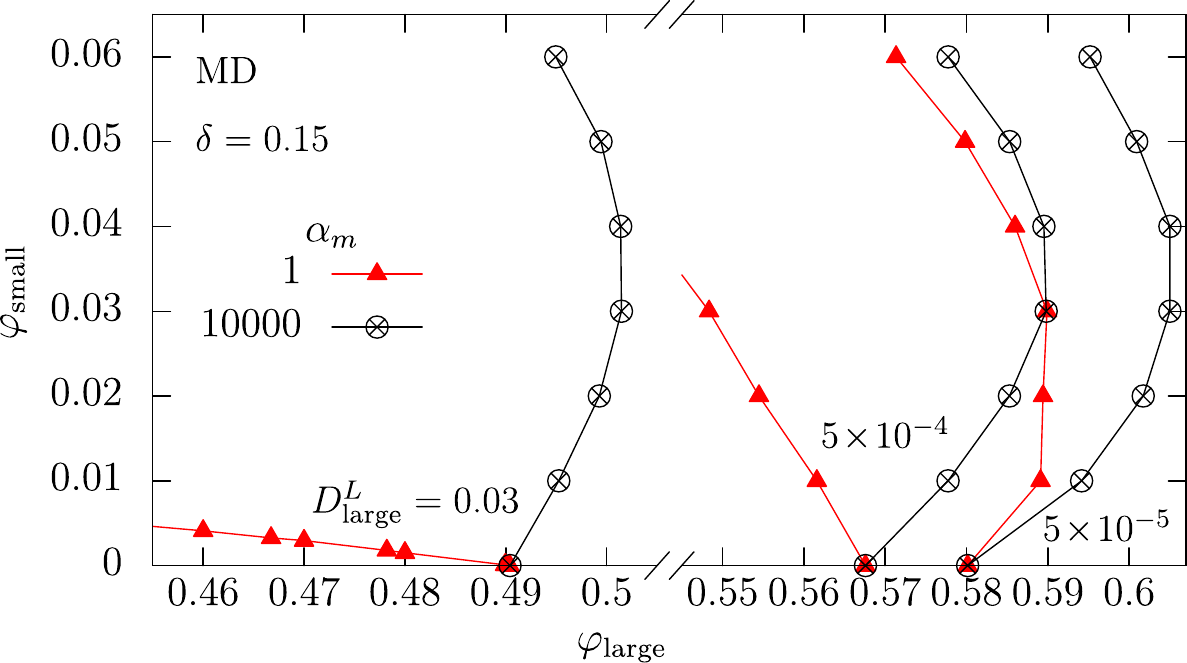}
\caption{\label{AO_mixtures}
  Iso-diffusivity lines for the long-time diffusion of large particles
  in an AO mixture with size ratio $\delta=0.15$ and mass ratio $\alpha_m=1$
  and $\alpha_m=10000$,
  in the state space spanned by $(\varphi_\text{large},\varphi_\text{small})$,
  for various values of $D^L_\text{large}$ as labeled.
}
\label{fig:aoisodiff}
\end{figure}

In the limit $\delta\to0$, one expects the fast (small) particles to remain
mobile even at densities where the slow (big) particles approach kinetic
arrest. The small particles may then be thought of as providing effective
entropically mediated depletion interactions between the large particles.
In colloid--polymer mixtures, small non-adsorbing polymers provide
short-range depletion attractions for the big colloidal particles,
and these cause
a reentrant melting of the big-particle glass upon addition
of small polymers \cite{Dawson.2001,Pham.2002,Zaccarelli:2006swsmct}.

The standard hard-sphere model for colloid--polymer mixtures is the
Asakura--Oosawa (AO) model \cite{Terentjev.2015,glass-mixtures:Zaccarelli.2004}. The model assumes that the small particles do not
interact among themselves, even though they keep excluded-volume interactions
with the big particles.
It was a striking finding reported from computer simulation
\cite{glass-mixtures:Zaccarelli.2004} that the reentry emerges from
the lines of constant diffusivity in the AO model
only if the small particles are sufficiently light.

Figure~\ref{fig:aoisodiff} shows iso-diffusivity lines for values of
$D^L_\text{large}$ close to the glass transition of the AO mixture
of Ref.~\cite{glass-mixtures:Zaccarelli.2004}. For $\varphi\lesssim0.58$,
the addition of lighter small particles causes
the structural-relaxation dynamics to speed up (iso-diffusivity lines bend
to larger $\varphi$), while the addition
of equally heavy small particles has the opposite effect.
This is consistent with earlier observations \cite{glass-mixtures:Zaccarelli.2004}.
However, extending these results to larger $\varphi$ that are closer to
the expected MCT glass transition ($\varphi\approx0.58$),
one notices that also the addition of heavy particles ($\alpha_m=1$)
leads to the reentrant enhancement of structural relaxation that was
observed for the addition of light particles ($\alpha_m=10000$).
Thus, although the pre-asymptotic approach to the MCT transition differs
for different $\alpha_m$ quite drastically, from extrapolation of our data
to $D^L_\text{large}=0$ one anticipates
that the kinetic transition itself does not depend
on the mass ratio.
%The one-component MCT gives the qualitatively correct
%dependence of the glass-transition line on the state parameters
%$(\varphi_\text{large},\varphi_\text{small})$, while two-component MCT
%does not \cite{glass-mixtures:Zaccarelli.2004}. We conclude that this
%signals a shortcoming of MCT in conjunction with the AO-mixture model,
%rather than the previously conjectured dependence of the transition on
%the mass ratio.

% Conclusion

In summary, we have shown that
increasing the mobility of fast particles in dynamically asymmetric mixtures
can induce a counter-intuitive slowing down of the
relaxation of the slower particles.
This regime is to be contradistincted from the one asymptotically close to,
but below the glass transition,
where an increase in mobility leads to faster relaxation that is
governed by universal behavior whose power laws are independent
of the kinetic parameters.

This surprising trend reversal calls for an interpretation.
The critical density $\varphi_c$ marks a change in transport mechanism.
Below $\varphi_c$, motion is liquid-like, and the hallmark of the
approach to the glass transition is a strong increase in viscosity,
respectively, the {friction},
$\zeta=(1/3k_\text{B}T)\int_0^\infty\langle\vec f(t)\cdot\vec f(0)\rangle\,dt$,
where $\vec f(t)$ is a fluctuating reduced force.
The diffusion then follows from the Einstein relation,
$D=k_\text{B}T/\zeta$.
Since increasing the mobility of the fast particles causes
the fluctuating forces to decay faster, this reduces the friction
and speeds up the diffusion.
Conversely, above $\varphi_c$, motion is solid-like, i.e., viscous
relaxation is ineffective. Here, {residual currents} determine the
diffusivity via the Green-Kubo relation,
$D=(1/3)\int_0^\infty \langle \vec v(t)\cdot\vec v(0)\rangle\,dt$, as an integral
over the velocity-autocorrelation function (VACF).
Consequently, an increase in mobility of the fast particles causes the VACF to
decay faster, and therefore, diffusion slows down.

Our interpretation is compatible with mode-coupling theory, which considers
the friction contribution to be dominant, but ignores residual-current
relaxation. For this reason, MCT can capture only the regime below $\varphi_c$,
but not the one above.
Studying systematically the influence of kinetic parameters therefore offers a clean way of identifying the
transition in transport mechanism without relying on power-law fits over
a limited range of relaxation times.

Both Brownian and Newtonian dynamics yield the same
qualitative results for both regimes.
This extends an earlier finding on the kinetic universality
\cite{glass-theory:Szamel.2004c} to
include size ratios and kinetic parameters.
%but universality for a different reason.
Therefore, it also encompasses the dynamics of mobile intruders in
crowded environments \cite{glass-theory:Sentjabrskaja.2016,glass-mixtures:Zaccarelli.2004},
with the caveat that one has to carefully separate the asymptotic
behavior from pre-asymptotic effects on the dynamics.

\begin{acknowledgments}
We thank A.~Meyer for discussions and helpful comments on the manuscript.
We gratefully acknowledge support by the DFG research unit FOR1394 ``Nonlinear Response to Probe Vitrification'',
and by the Austrian Science Fund (FWF) under grant I~2887-N27 (TF).
\end{acknowledgments}

\bibliography{lit}
\bibliographystyle{apsrev4-1}

\end{document}